\documentstyle[emulateapj,psfig,apjfonts]{article}
\def\sax{SAX~J1808.4--3658}
\def\etal{{\rm et~al.\ }}

\lefthead{GAENSLER, STAPPERS AND GETTS}
\righthead{TRANSIENT RADIO EMISSION FROM \sax}

\begin{document}
\title{Transient radio emission from SAX~J1808.4--3658}
\submitted{(To appear in {\em The Astrophysical Journal Letters})}
\author{B. M. Gaensler\altaffilmark{1,5}, B. W. Stappers\altaffilmark{2}
and T. J. Getts\altaffilmark{3,4}}

\altaffiltext{1}
{Center for Space Research, Massachusetts Institute of Technology,
70 Vassar Street 37--667, Cambridge, MA 02139, USA; email: bmg@space.mit.edu}
\altaffiltext{2}{Astronomical Institute ``Anton Pannekoek'', Kruislaan 403,
1098 SJ Amsterdam, The Netherlands; email: bws@astro.uva.nl}
\altaffiltext{3}{Australia Telescope National Facility, CSIRO, 
PO Box 76, Epping NSW 1718, Australia; email: tracy.getts@atnf.csiro.au}
\altaffiltext{4}{School of Mathematics, Physics, Computing and Electronics,
Macquarie University, NSW 2109, Australia}
\altaffiltext{5}{Hubble Fellow}

\begin{abstract}
We report on the detection of radio emission from the accretion-powered
X-ray millisecond pulsar SAX J1808.4--3658, using the Australia
Telescope Compact Array.  We detected a $\sim$0.8~mJy source at the
position of SAX~J1808.4--3658 on 1998 April 27, approximately one day
after the onset of a rapid decline in the X-ray flux; no such source
was seen on the previous day.  We consider this emission to be related
to the radio emission from other X-ray binaries, and is most likely
associated with an ejection of material from the system.  No radio
emission was detected at later epochs, indicating that if
SAX~J1808.4--3658 is a radio pulsar during X-ray quiescence then its
monochromatic luminosity must be less than $L_{\rm 1.4\,GHz} \sim
6$~mJy~kpc$^2$.
\end{abstract}


%

\section{Introduction}

Millisecond pulsars (MSPs) have long been thought to be the endpoint in the
evolution of low-mass X-ray binaries (LMXBs) (\cite{acrs82};
\cite{rs82}). Although
the link between the LMXBs and MSPs is strong, evidence that LMXBs did
indeed contain objects spinning at millisecond periods was, until
recently, still
missing. The quasi-periodic oscillations, especially near 1\,kHz
(\cite{vsz+96}), recently discovered in X-ray binary systems offer
indirect evidence for the existence of weakly magnetized neutron stars
with millisecond periods. However,
the recent discovery of 2.49~ms coherent X-ray pulsations from the LMXB
SAX~J1808.4--3658 (\cite{wv98}; \cite{cm98})
now gives strong support to the picture
that LMXBs are the progenitors of MSPs; this system
provides the best evidence yet that a low-field neutron star can be spun up
to millisecond periods via accretion from its companion.

There are a number of processes which could potentially produce radio
emission from such a system. An exciting possibility is that \sax\
will, at some point, turn on as a radio pulsar, producing
pulsed radio emission characteristic of MSPs. Timing of such pulses could
allow an improved determination of the astrometric, rotational and orbital
parameters of the system as well as the determination of possible
post-keplerian and tidal effects. Furthermore they could probe any wind
produced by the interaction between the pulsar and its companion such as in
the eclipsing binary millisecond pulsar systems (e.g. \cite{fbb+90}).
Such emission could potentially be heavily
scattered by this interaction or by interactions with material excreted from
the system (e.g. \cite{rst89}). Thus it would
appear unpulsed and could only be detected as a continuum
source. Alternatively, unpulsed radio emission could be produced by the
interaction of the relativistic pulsar wind with the interstellar medium
(e.g. \cite{fggd96}).  Finally, radio emission could be
associated with the accretion process and X-ray outburst, as seen in a
significant fraction of the X-ray binary population (\cite{hh95}; \cite{fh99}).

We here report on a search for radio emission during the
outburst and then also during quiescence, aimed at 
testing these possibilities. In \S\ref{obs} we describe our
observations and analysis, while in \S\ref{results} we demonstrate the
detection of radio emission from the system, and discuss its implications.

\section{Observations and Reduction}
\label{obs}

All observations were made with the Australia Telescope Compact Array (ATCA;
\cite{fbw92}), an east-west synthesis array
located near Narrabri, NSW, Australia; these observations are summarized in
Table~1.  The ATCA can observe two frequencies simultaneously;
for all epochs except 1998 Nov 30, we alternated, approximately every 20
minutes, between observing with a 1.4/2.5~GHz combination and a 4.8/8.6~GHz
combination. On 1998 Nov 30, observations were only made at
1.4/2.5~GHz.  A bandwidth of 128~MHz was used at each frequency.  A pointing
center RA (J2000) $18^{\rm h}08^{\rm m}13^{\rm s}$, Dec (J2000)
$-36\arcdeg57\arcmin18\arcsec$ was observed in all cases, approximately $3'$
from the position of SAX~J1808.4--3658\ (\cite{ghg99}). Amplitudes were
calibrated using observations of PKS~B1934--638, assuming a flux density for
this source of 15.0, 11.1, 5.8 and 2.8~Jy at 1.4, 2.5, 4.8 and 8.6~GHz
respectively (where 1~Jy~$=10^{-26}$~W~m$^{-2}$~Hz$^{-1}$). Phases were
calibrated using observations once per hour of PKS~B1934--638 (at 1.4 \& 2.5
GHz) and PMN~J1733--3722 (at 4.8 \& 8.6~GHz).

Data were reduced in the {\tt MIRIAD} package using standard techniques.
For each frequency in each observation, images were formed using natural
weighting and excluding baselines shorter than 1.5~km.  Sidelobes around
detected sources were deconvolved using the {\tt CLEAN} algorithm. Each
image was then smoothed to the appropriate diffraction limit, and
corrected for the mean primary beam response of the ATCA antennas.

\section{Results and Discussion}
\label{results}

The position for SAX~J1808.4--3658, as determined by observations of its optical
counterpart, is at RA (J2000) $18^{\rm h}08^{\rm m}27\farcs54$, Dec (J2000)
$-36\arcdeg58\arcmin44\farcs3$ with uncertainties
of $\sim0\farcs8$ in each coordinate (\cite{ghg99}). In all observations
except Obs~1, the only radio source detected in a $\sim5\arcmin$ region surrounding this
position was a double-lobed radio galaxy, NVSS~180824--365813 (\cite{ccg+98});
no source was seen at the position of \sax.
The resolution and limiting sensitivity for these non-detections
are summarized in Table~1.

\begin{figure*}
\centerline{\psfig{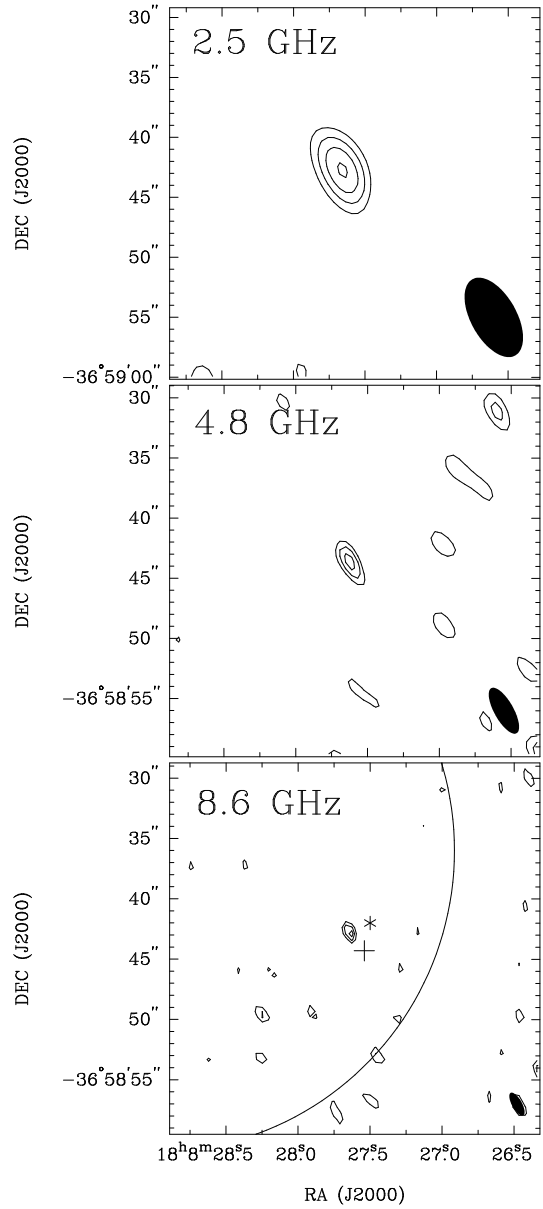}}
\caption{2.5, 4.8 and 8.6~GHz images of SAX~J1808.4--3658\ on 1998 Apr
27. Contours in each image are at the levels of 0.3, 0.45, 0.6 and
0.75~mJy~beam$^{-1}$, while the synthesized beam (FWHM) is shown at lower
right of each panel. In the 8.6~GHz image, a cross and a star mark the
optical positions for \sax\
of Giles et al. (1999) and Roche et al. (1998) respectively,
while the large arc corresponds to the edge of the X-ray timing error circle
of Chakrabarty \& Morgan (1998).}
\label{fig_contours}
\end{figure*}

In Obs~1 an unresolved radio source near the position of \sax\
was detected at the
$\sim4\sigma$ level at each of 2.5, 4.8 and 8.6~GHz, as shown in
Figure~\ref{fig_contours}; flux densities are given in
Table~2. The source was not detected at 1.4 GHz. 
We attemped various approaches to the imaging, deconvolution and 
fitting processes. The results of these suggest a systematic
uncertainty in the flux densities for the source of $\sim$50\%, 
a value to
be expected when deconvolving a weak source under conditions
of poor $u-v$ coverage. Our best position for this source is RA (J2000)
$18^{\rm h}08^{\rm m}27\farcs6$, Dec (J2000)
$-36\arcdeg58\arcmin43\farcs9$ with an uncertainty of $\sim0\farcs5$ in
each coordinate. In the bottom
panel of Figure~\ref{fig_contours}, our position and the optically
determined positions of Giles et al. (1999)\nocite{ghg99} and Roche et
al.  (1998)\nocite{rcm+98} are compared.  All three positions are
consistent within the quoted uncertainties.

The source shown in Figure~\ref{fig_contours} is almost certainly not an
artifact. It is not at the phase center, and was
detected at three different frequencies
and for a variety of different weighting schemes and
combinations of baselines. While the probability of finding an unrelated
radio source within a few arcsec of SAX~J1808.4--3658\ is low in any case,
we note that this source was not detected at any other epoch, despite the
improved sensitivity of these later observations. Furthermore, the region
was observed with the Very Large Array (VLA) on 1998 Apr 26, the day
immediately before Obs~1, and no source was detected at this position down
to a comparable sensitivity (R.M. Hjellming, private communication). 
Therefore, from its transient nature and positional
coincidence with the optical counterpart of SAX~J1808.4--3658, we conclude
that this radio source is associated with the system.

Our data lack the time-resolution required to search for pulsed radio
emission from this source. However, X-ray emission due to accretion was
present at the time of our detection, and it is likely that if any radio
emission was being produced in the pulsar magnetosphere it would be quenched
by this process. Thus the source we have detected is most likely not
emission associated with the magnetosphere. It is also unlikely that the
source corresponds to emission from a pulsar wind (if such a wind even
exists at this point in the system's evolution) --- if we assume that the
disappearance of the source between Obs 1 and Obs 2 is due to synchrotron
losses, then this cooling time-scale implies a magnetic field in the wind of
$\ga$2~G, many orders of magnitudes higher than observed for other pulsars
(e.g.\ \cite{msk93}; \cite{fggd96}).
The detection of radio
emission only at an epoch during which X-rays were still being generated
provides strong evidence that this radio source is related to the accretion
process.

Radio emission has been detected from $\sim$25\% of all X-ray binaries
(\cite{fh99}).  In cases for which this emission has been resolved, it
takes the form of jets being emitted from the system, often at
relativistic velocities (\cite{fbw97}). The burst properties and rapid
X-ray variability of SAX~J1808.4--3658 suggest that it is an
atoll source, i.e.\ a low magnetic field neutron star accreting at
about 10\% of the Eddington limit (\cite{wv98a}).
Fender \& Hendry (1999) review the
radio properties of different types of persistent 
X-ray binaries, and show that
most atoll sources show no radio emission --- the only detections have
been transient, and at the mJy level.  When transient radio emission is
seen from the atoll
sources, the spectrum is first seen absorbed, but then becomes optically
thin and takes on a synchrotron spectrum with $\alpha \approx -0.5$
($S_\nu \propto \nu^{\alpha}$). The emission then decays away through
adiabatic losses (e.g. \cite{hh95}).

The properties of the radio emission seen here for SAX J1808.4--3658 are
consistent with this behavior.  While the spectrum of the source is very
poorly constrained by our data, the flux
densities for it at 2.5~GHz and above are
consistent with $\alpha \approx -0.5$, while the non-detection at 1.4~GHz is
suggestive of a low-frequency turnover probably due to
self-absorption in the ejecta.

We note that on 1998 Apr 26, the day before our radio detection, the
X-ray flux suddenly deviated from exponential decay and began to
rapidly decrease (\cite{grsc98}). As discussed by these authors, there
are two different mechanisms which can produce such a cut-off, either
the onset of the so-called ``propeller phase'' or an instability of the
accretion disk. In both cases the abrupt change corresponds to ejection
of material from the system. The appearance of transient radio emission
just after this event suggests that it is this ejection of material
which has produced the source we see here.  If we assume that emission
above 2.5~GHz corresponds to optically thin, incoherent, unbeamed,
synchrotron emission, the $10^{12}$~K Compton limit on the brightness
temperature corresponds to a minimum scale for the emission of
$\sim$12~light~seconds at a distance of 4~kpc, much larger than the
$\sim$1~light second binary separation of the orbit (\cite{cm98}).
Thus the emission is coming from well beyond the binary, as would be
expected if it has resulted from an expulsion of material from the
system.

At later times, no radio emission was detected at the position of
SAX~J1808.4--3658. This is not necessarily indicative that the radio
pulse mechanism is not yet functioning. Despite strong limits on the
flux density from our non-detections, the apparently large distance to
the system of $\sim$4\,kpc corresponds to a 3$\sigma$ monochromatic
luminosity limit of $L_{\rm 1.4\,GHz} \sim 6$~mJy~kpc$^2$, greater than
that of the majority of known MSPs (\cite{tmlc95}). Furthermore, Ergma
et al. (1999)\nocite{ea99} have proposed that this system may only be
detected at shorter wavelengths ($\lambda < 3$\,cm) due to free-free
absorption by material excreted from the system.

\section{Conclusion}

We have detected a transient radio source coincident with
SAX~J1808.4--3658; the source turned on within a day, and then had
disappeared again a month later.  We interpret this source as radio
emission associated with ejection of material, as seen in other X-ray
binaries, and in this case possibly associated with the onset of a
propellor phase or a disk instability the day before the source was
detected.  The spectrum and light curve of this source are essentially
unconstrained, however, and the source should certainly be searched for
radio emission next time it is in outburst. Eventually, it is hoped
that this source will emerge as a radio MSP; radio searches in
quiescence should continue to be carried out in anticipation of this.

\begin{acknowledgements}

We thank Rob Fender and Deepto Chakrabarty for useful discussions, Bob
Hjellming for communicating the results of his VLA observations, and Scott
Cunningham, Lucyna Kedziora-Chudczer, Robin Wark and Mark Wieringa for
assistance with the observations.  The Australia Telescope is funded by the
Commonwealth of Australia for operation as a National Facility managed by
CSIRO.  BMG acknowledges the support of NASA through Hubble Fellowship grant
HF-01107.01-98A awarded by the Space Telescope Science Institute, which is
operated by the Association of Universities for Research in Astronomy, Inc.,
for NASA under contract NAS 5--26555. BWS is supported by NWO Spinoza
grant 08-0 to E.P.J. van den Heuvel.

\end{acknowledgements}



\clearpage

\clearpage

\begin{table*}
\begin{center}
TABLE~1

{\sc ATCA observations of SAX~J1808.4--3658.\label{tab_limits}}

\vspace{1mm}
\begin{tabular}{cccccc}  \hline \hline
Obs    & Date &  Time on & Frequency   & Resolution   & rms \\
       &      & Source (h) & (GHz)     & ($''$)   & (mJy~beam$^{-1}$) \\ \hline
1      & 1998 Apr 27  & 4.5 & 1.4    & $13.0\times7.2$  & 0.22 \\
      &  &           &        2.5    &  $7.3\times3.8$ & 0.21  \\
      & &           &    4.8    & $4.3\times1.6$ & 0.15 \\
      & & & 8.6    & $2.1\times0.8$ & 0.18  \\
2      & 1998 May 28, & 11 &   1.4 & $11.3 \times 6.6$ & 0.42 \\
       & 1998 May 30, & &  2.5  & $6.2 \times3.5$ & 0.19 \\
       & 1998 Jun 01  & &   4.8   & $3.2\times1.6$ & 0.11 \\
       &    &&  8.6    & $1.5 \times 0.8$ & 0.12 \\
3      & 1998 Jun 10 & 2.5 &    1.4   &  $13.6\times6.0$  & 0.81 \\
       &    & & 2.5    & $7.6\times3.1$  & 0.32 \\
       &    & & 4.8    & $4.4\times1.3$  & 0.19 \\
       &    & & 8.6    &  $2.1\times0.7$   & 0.20 \\
4      & 1998 Oct 05 & 9 &    1.4   &  $13.0\times8.2$ & 0.19 \\
       &    & & 2.5    & $7.2 \times4.6$  & 0.16 \\
       &    & & 4.8    & $3.7\times2.5$        & 0.07 \\
       &    & & 8.6    & $2.1\times1.7$      & 0.10 \\
5      & 1998 Nov 30 & 6.5 &    1.4   &  $13.5\times7.2$  & 0.12 \\ 
       &    & & 2.5   & $7.5\times4.0$  & 0.10 \\ \hline
\end{tabular}
\end{center}
\end{table*}

\begin{table*}
\begin{center}
TABLE~2

{\sc Results of Observation 1.\label{tab_detect}}

\vspace{1mm}
\begin{tabular}{cccc}  \hline \hline
  Frequency     & Flux of source  \\
    (GHz)      & (mJy) \\ \hline
      1.4    & $\ldots$ \\
      2.5    & 0.8 \\
      4.8    & 0.8 \\
      8.6    & 0.8 \\ \hline
\end{tabular}
\end{center}
\end{table*}

\end{document}